\documentclass[11pt]{article}
\usepackage{hyperref}
\pdfoutput=1
\begin{document}
\title{Homage to Bob Brodkey at 85:  Ejections, Sweeps and Reynolds Shear Stress Generation}
\author{James M. Wallace and James H. Duncan\\
\\Department of Mechanical Engineering\\ 
and the Burgers Program for Fluid Dynamics\\ 
University of Maryland, College Park, MD 20742, USA}
\maketitle
\begin{abstract}
Almost 50 years ago Bob Brodkey and his student, Corino, conceived of and carried out a visualization experiment for the very near wall region of a turbulent pipe flow that, together with the turbulent boundary layer visualization of Kline et al., excited the turbulence research community.  Using a high speed movie camera mounted on a lathe bed that recorded magnified images in a frame of reference moving with the flow, they observed the motions of sub-micron particles in the sub-, buffer- and log-layers.  Surprisingly, these motions were not nearly so locally random as was the general view of turbulence at the time.  Rather, connected regions of the near wall flow decelerated and then erupted away from the wall in what they called "ejections".  These decelerated motions were followed by larger scale connected motions toward the wall from above that they called "sweeps".   They estimated that ejections accounted for $70\%$ of the Reynolds shear stress at $Re_d = 20,000$ while only occurring about $18\%$ of the time. This fluid dynamics video shows short sequences, at three Reynolds numbers, from the original Corino and Brodkey visual study.  In the following decades it inspired numerous laboratory and simulation studies aimed at unraveling the structure of bounded turbulent flows and understanding the turbulent transport processes within them.
 \end{abstract}
 
\section{Introduction}
 This video is made up of three sequences from the original film of a turbulent pipe flow, entitled "The Wall Region in Turbulent Flow", that formed the basis of the 1965 Ohio State University dissertation of E. R. Corino.  Subsequently he and his dissertation advisor, Bob Brodkey, published a detailed set of observations from the film [J. Fluid Mech. {\bf 37}, 1- 30 (1969)].  The film was remarkable because the experimental setup allowed the authors to view, in a frame of reference moving with the flow, sub-micron particle motions in the near wall region at a succession of different Reynolds numbers.  They achieved this using a high speed movie camera mounted on a lathe bed that translated with and recorded magnified and time resolved images of the sub-layer, buffer-layer and lower part of the log-layer in the pipe flow.  Surprisingly, these particle motions were not nearly so locally random as was the general view of turbulence at the time.  Rather, connected regions of the near wall flow decelerated and then erupted away from the wall in what they called "ejections".  These decelerated motions were followed by larger scale connected motions toward the wall from above that they called "sweeps".  In the following decades this visualization study, along with that of Kline et al. [J. Fluid Mech. {\bf 30}, 741 - 773 (1967)] showing "low-speed streaks" and "bursting" in the wall region of a boundary layer, inspired numerous laboratory and simulation studies aimed at unraveling the structure of bounded turbulent flows and understanding the transport processes within them.  
 
 For example, Wallace, Brodkey and Eckelmann [J. Fluid Mech. {\bf  54}, 39 - 48 (1972)] attempted to quantify the visual observations of Corino and Brodkey by conceiving of and carrying out a quadrant analysis of the Reynolds shear stress in a turbulent oil channel flow.  This analysis exploited the information contained in the signs of the velocity fluctuations by classifying the streamwise and wall normal fluctuations into four quadrants based on their pairs of signs, and calculated the contribution of each quadrant to the Reynolds shear stress.  Quadrants 2 (-u, +v) and 4 (+u and -v) roughly correspond to the "ejections" and "sweeps" of the Corino and Brodkey visual observations.  Of course, other pairs of variables that constitute covariances can be treated with quadrant analysis.
 
 \section{Description of the Video}
 
 In this video, the setup of the Corino and Brodkey visual experiment is first illustrated. The flowing fluid was trichloroethylene with a refractive index that closely matched that of the glass pipe.  The pipe itself was submerged in still trichloroethylene, contained in a glass rectangular channel, making it possible to view the moving particles through the walls of the channel and pipe with very little refraction.  The  remaining slight refraction reveals the pipe wall which appears as a dark horizontal line at the bottom of the  video moving to the right in the translating  frame of reference. The three-dimensional field of view extends out from the wall to $y^+ = 45$ at $Re_d = 20,000$ and to $y^+ \approx 75$ at $Re_d = 40,000$.  The depth of the field of view for these two Reynolds numbers is $z^+ \approx 18$ and $z^+ \approx 29$ respectively.  The video sequences are for three Reynolds numbers based on pipe diameter.  
 
 The striking observation at $Re_d = 20,000$ is how organized the flow appears.  Groups of particles decelerate together, more or less parallel to the wall, and then eject outward at large angles (in the moving frame of reference) relative to the wall.  These ejection events are frequently followed by groupings of particles, extending over larger regions, that sweep in from above with higher velocities than that nearer the wall.  It is also frequently observed that particles move past each other, indicating velocity variation and shear normal to the viewing plane.  When the Reynolds number is increased to 30,000 where the depth of the field of view is larger, the ejection and sweep events often appear to involve the same particles in swirling motions of what seem to be quasi-streamwise vortices.  These vortices increase in frequency and intensity when the Reynolds number is increased to 40,000.  Corino and Brodkey estimated that ejections accounted for $70\%$ of the Reynolds shear stress at $Re_d = 20,000$, while only occurring about $18\%$ of the time, demonstrating their dynamical importance.


%
\end{document}